\documentclass[runningheads,a4paper]{cmmr2023}
\usepackage{times}
\usepackage{amssymb}
\usepackage{graphicx}
\usepackage{listings}
\usepackage{xcolor}
\usepackage{subcaption}
\usepackage[hyphens]{url}
\usepackage{hyperref}
\usepackage{multirow}
\newcommand{\keywords}[1]{\par\addvspace\baselineskip
\noindent\keywordname\enspace\ignorespaces#1}

\definecolor{codegreen}{rgb}{0,0.6,0}
\definecolor{codelightgreen}{rgb}{0,0.6,0.4}
\definecolor{codelightblue}{rgb}{0,0.5,0.7}
\definecolor{codegray}{rgb}{0.5,0.5,0.5}
\definecolor{codepurple}{rgb}{0.89,0,0.62}
\definecolor{codered}{rgb}{0.69,0,0.20}
\definecolor{codeblue}{rgb}{0.4,0,0.6}
\definecolor{backcolour}{rgb}{0.95,0.95,0.97}
\definecolor{lightblue}{rgb}{0.0,0.0,0.9}
\definecolor{mauve}{rgb}{0.58,0,0.82}

\lstdefinelanguage{XQuery}{
    keywords={},
    keywords=[2]{
        for,let,return,where,declare,default,function,element,variable,external,
        copy,modify,replace,delete,insert,node,nodes,first,module,namespace,import
    },
    moredelim=*[s][\color{codelightgreen}]{<}{>},
    alsoletter={-},
    keywords=[3]{
        in,eq,and,not,as,with,into,at, doc, distinct-values
    },
    keywordsprefix=\$,
    morestring=[b]',
    morestring=[b]",
    morecomment=[n]{(:}{:)}
}
\lstdefinestyle{mystyle}{
  backgroundcolor=\color{backcolour},
  commentstyle=\color{codegreen},
  keywordstyle=\color{codelightblue},
  keywordstyle=[2]\color{codered},
  keywordstyle=[3]\color{codeblue},
  numberstyle=\tiny\color{codegray},
  stringstyle=\color{codepurple},
  basicstyle=\ttfamily\scriptsize,
  breakatwhitespace=false,         
  breaklines=true,                 
  captionpos=b,                    
  keepspaces=true,                 
  numbers=left,                    
  numbersep=5pt,                  
  showspaces=false,                
  showstringspaces=false,
  showtabs=false,                  
  tabsize=2
}

\lstdefinelanguage{XML}
{
  morestring=[s][\color{mauve}]{"}{"},
  morestring=[s][\color{black}]{>}{<},
  morecomment=[s]{<?}{?>},
  morecomment=[s][\color{dkgreen}]{<!--}{-->},
  stringstyle=\color{black},
  identifierstyle=\color{lightblue},
  keywordstyle=\color{red},
  morekeywords={xmlns,xsi,noNamespaceSchemaLocation,type,id,x,y,source,target,version,tool,transRef,roleRef,objective,eventually}
}

\begin{document}

\mainmatter  

\title{SANGEET: A XML based Open Dataset for Research in Hindustani Sangeet}

\titlerunning{SANGEET: A XML based Open Dataset for Research in Hindustani Sangeet}

%
%
\author{Chandan Misra\and Swarup Chattopadhyay \thanks{We would like to thank the undergraduate students of School of Computer Science \& Engineering for helping create the dataset.}}
%
\authorrunning{Chandan Misra and Swarup Chattopadhyay}

\institute{School of Computer Science \& Engineering, XIM University\\ \email{chandan@xim.edu.in, swarupc@xim.edu.in}}

%

\maketitle

\begin{abstract}
It is very important to access a rich music dataset that is useful in a wide variety of applications. Currently, available datasets are mostly focused on storing vocal or instrumental recording data and ignoring the requirement of its visual representation and retrieval. This paper attempts to build an XML-based public dataset, called SANGEET, that stores comprehensive information of \textit{Hindustani Sangeet} (North Indian Classical Music) compositions written by famous musicologist \textit{Pt. Vishnu Narayan Bhatkhande}. SANGEET preserves all the required information of any given composition including metadata, structural, notational, rhythmic, and melodic information in a standardized way for easy and efficient storage and extraction of musical information. The dataset is intended to provide the ground truth information for music information research tasks, thereby supporting several data-driven analysis from a machine learning perspective. We present the usefulness of the dataset by demonstrating its application on music information retrieval using XQuery, visualization through \textit{Omenad} rendering system. Finally, we propose approaches to transform the dataset for performing statistical and machine learning tasks for a better understanding of \textit{Hindustani Sangeet}. The dataset can be found at \url{https://github.com/cmisra/Sangeet}.

\end{abstract}

\section{Introduction}
Having access to free, well-maintained databases of music is a crucial resource for researchers. In the case of Indian Classical Music, this is also true since it has been shown to be important for high-quality research in music information retrieval (MIR) \cite{chithra2015music,kirthika2012review,sridhar2009raga,murthy2018content} and musicological analysis using machine learning \cite{ujlambkar2012mood,sridharan2018similarity,patel2017raag}, deep learning \cite{madhusudhan2019deepsrgm,sharma2021classification,nag2022application,pendyala2022towards}, etc. Several high-quality datasets, \cite{compmusic} and \cite{dunya} for example, for research in MIR and computational musicology can be found in the published literature.

Although audio recording-based music corpora are essential in certain types of music applications,  studies of existing literature reveal a dearth of substantial research related to the varied domains of \textit{Sound and Music Computing} (SMC), especially in the design and development of interfaces for expressing Indic Music on an electronic medium. One of the domains that interest us is the creation of an Indian music environment through the transcription and rendering of an Indic music piece using Indic notation systems and Indic language script. The ability to compose music electronically entirely in an Indian music environment necessitates the emergence of research in different domains in SMC.  This requires a musicological analysis of the grammar and structure of the music sheets presently in use and the consequent development of musical fonts and rendering engines (\cite{lilypond} for staff notation for example). Needless to say, such endeavors would motivate the research community to create models for Indic music notation systems and their language bases \cite{misra2016swaralipi} and provide ample opportunities to work in building interfaces for music expressions on an electronic medium.

The mere enabling of music practitioners in composing music electronically is insufficient unless we have tools to exchange such musical information seamlessly across applications. Consequently, this establishes the need for the development of a common music exchange format to communicate music independent of any genre, notation system, language script, and music sheet structure \cite{misra2021sangeetxml}. XML-based formats for exchanging musical information have existed for quite some time \cite{misra2021sangeetxml,good2001musicxml,baggi2009ieee} and are being adopted by extremely robust and popular notation software like \textit{Finale} \cite{finale}, \textit{Sibelius} \cite{sibelius}, \textit{MuseScore} etc. Additionally, the ability to store musical information in XML solves the problem of archiving our historic musical art form as an electronic database.

One of the most authentic sources of Hindustani Sangeet is the compositions published in the book \textit{Hindusthani Sangeet Paddhati-Kramik Pustak Malika} which comprises of approximately $1900$ compositions belonging to North Indian Classical Music penned by \textit{Pt. Vishnu Narayan Bhatkhande} ($1860$ - $1936$). In order to reach a greater number of music students and scholars, the first volume of \textit{Kramik Pustak Malika} has been translated to \textit{Hindi} language in $1953$ by prominent music scholar \textit{Dr. Laxminarayan Garg}. This paper introduces SANGEET, arguably the first XML-based music corpora that try to capture comprehensive musical information contained in these rich music sources to apply in various music applications like music transcription, visualization, MIR, computational musicology, etc. We begin the preparation of the dataset with the second volume of \textit{Kramik Pustal Malika} book series and our objective is to store compositions of different genres in a carefully crafted XML database to preserve comprehensive musical information in a single format. This will provide the users to obtain a standard framework for efficient and easy access to the dataset that can be easily transformed to apply to various applications. We refer to three music applications related to visualization, MIR, and machine learning in support of the coverage, quality, and accessibility of SANGEET.

\section{The Organization and Access of SANGEET}
\label{sec:dataset-creation-description}
Pt. Vishnu Narayan Bhatkhande is the pioneer for providing a comprehensive theoretical foundation of Hindustani Sangeet in a published form in his six-volume book series titled \textit{Hindustani Sangeet Paddhati, Kramik Pustak Malika} written in \textit{Marathi} language in $1920$. His book contains a comprehensive description of music symbols for realizing musical components including notes (Svar), time signatures (Lay), beats (Taal), ornaments (Alankar) etc. The dataset created in the current work has been taken from the Hindi translation of the second volume of the series. The second volume of the book series contains a total of $319$ compositions belonging to $10$ different raags. The present work takes these written compositions as a source of musical information to create the database for Hindustani Sangeet to be used in various applications.


We have taken $116$ compositions of the three highest frequent raags i.e. raag \textit{Bhairav} ($42$), \textit{Todi} ($39$), and \textit{Poorvi} ($35$) respectively, from the entire collection of $319$ compositions for performing our experimental analysis. Eventually, the entire collection of compositions from all six volumes will be preserved in the dataset for applications related to music information retrieval, music-sheet visualization, etc.
 

The dataset consists of a number of XML documents that is equal to the number of compositions in the dataset i.e. each XML document represents a single composition of the dataset. The XML documents are equipped with meaningful tags to store all the necessary musical information for the compositions. The format of the XML files is validated against a schema definition document so that the format of the dataset or compositions are preserved. The schema definition document is an XML Schema Definition (XSD) file against which each XML document is checked and validated for legal elements and attributes. The XSD consists of four parts namely \textit{info}, \textit{taal}, \textit{raag}, and \textit{sheet} responsible for storing metadata, rhythmic, melodic, structural, and notational information in the XML files.


The metadata linked to the musical composition is represented by the \textit{info} portion. It contains information on the catalog, the genre, and the notational system as shown in Listing \ref{lst:info} describing the first composition of the second volume of the book.

\begin{lstlisting}[language=XML, caption=Info Part of XML file depicting metadata, label=lst:info]
<INFO>
	<TITLE>Composition 1 Volume 2 Kramik Pustak Malika</TITLE>
	<AUTHOR>Pt. Vishnu Narayan Bhatkhande</AUTHOR>
	<NOTATION_SYSTEM>Bhatkhande</NOTATION_SYSTEM>
	<DATE_TIME>1923</DATE_TIME>
	<GENRE>Hindusthani Sangeet</GENRE>
	<ADDITIONAL>
		<ENTRY>http://ndl.iitkgp.ac.in/document/R2pPWGRxdkRWWnlvOVdPYzdzaWpTV0pYYTFIT0VnNTB6V1dnR1dJVW1kUT0</ENTRY>
	</ADDITIONAL>
</INFO>
\end{lstlisting}

The rhythmic foundation of Indian music is provided by \textit{taal}. Indic music has nearly hundreds of Taals, each with its own specific composition that includes the name, \textit{Bibhaga} or measure, \textit{Maatra} or the number of beats, \textit{Avartana} or the number of cycles per line, etc. Additionally, Taal has two designated beat indices, known as \textit{Taali} and \textit{Khali}, to signify stressed or unstressed strokes in addition to a specific beat pattern to uniquely identify a Taal. These patterns, which are required to portray the Taal graphically or as a music sheet accompanied by an Indian percussionist, have been illustrated as a series of numbers (seen in Listing \ref{lst:taal}). Additionally, the regular expression specifies the expression for a beat pattern, making it easier to query the Taal's structure.

\begin{lstlisting}[language=XML, caption=Taal Part of XML file depicting Taal and its sub-components, label=lst:taal]
<TAAL>
	<TAAL_NAME>Tritaal</TAAL_NAME>
	<BIBHAGA>4</BIBHAGA>
	<MAATRA>16</MAATRA>
	<AVARTANA>1</AVARTANA>
	<BEAT_PATTERN>4-4-4-4</BEAT_PATTERN>
	<ALTERNATE_BEAT_PATTERN>NA</ALTERNATE_BEAT_PATTERN>
	<TAALI_COUNT>3</TAALI_COUNT>
	<KHALI_COUNT>1</KHALI_COUNT>
	<TAALI_INDEX>1-5-13</TAALI_INDEX>
	<KHALI_INDEX>9</KHALI_INDEX>
</TAAL>
\end{lstlisting}

Raag provides the melodic framework to Hindustani Sangeet and each raag can be identified by characteristics like \textit{Arohana} and \textit{Avarohana}, which are ascending or descending movements made up of a series of notes, \textit{Vadi} and \textit{Samvadi}, which are consonant and dissonant notes, and classification forms like \textit{Pakad} and \textit{Jaati}. These characteristics are note sequences and have been encoded using \textit{Ome Swarlipi} \cite{omeswarlipi}, the same rendition we use for storing notes in our dataset.


\begin{lstlisting}[language=XML, caption=Raag Part of XML file depicting Raag and its sub-components, label=lst:raag]
<RAAG>
	<RAAG_NAME>Yaman</RAAG_NAME>
	<THAAT>Kalyan</THAAT>
	<AROHANA>n-r-g-M-d-n-su</AROHANA>
	<AVAROHANA>su-n-d-p-M-g-r-s</AVAROHANA>
	<VADI>g</VADI>
	<SAMVADI>n</SAMVADI>
	<JAATI>Sampoorna</JAATI>
	<PAKAD>nlrgr-s-pMg-su</PAKAD>
</RAAG>
\end{lstlisting}

\begin{figure}[!ht]
    \centering
    \begin{subfigure}{\textwidth}
    \centering
    \includegraphics[scale=0.3]{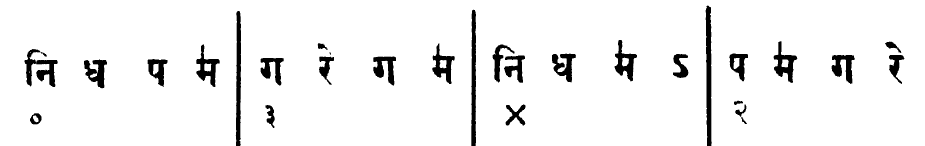}
    \caption{}
    \label{fig:sheet}
    \end{subfigure} \\
    \begin{subfigure}{0.48\textwidth}
\begin{lstlisting}[language=xml, label=lst:song-1]
<SHEET>
	<TOTAL_LINE></TOTAL_LINE>
	<LINES>
		<LINE INDEX="1">
			<ROW INDEX="1">
				<COL INDEX="1">
					<NOTE_COUNT>1</NOTE_COUNT>
					<CONTENT>n</CONTENT>
				</COL>
				<COL INDEX="2">
					<NOTE_COUNT>1</NOTE_COUNT>
					<CONTENT>d</CONTENT>
				</COL>
				<COL INDEX="3">
					<NOTE_COUNT>1</NOTE_COUNT>
					<CONTENT>p</CONTENT>
				</COL>
				<COL INDEX="4">
					<NOTE_COUNT>1</NOTE_COUNT>
					<CONTENT>M</CONTENT>
				</COL>
			</ROW>
			......
\end{lstlisting}
    \caption{}
    \label{fig:part-1}
    \end{subfigure}
    \hfill
    \begin{subfigure}{0.48\textwidth}
\begin{lstlisting}[language=xml, label=lst:song-2]
<LINE INDEX="2">	
			<ROW INDEX="1">
				<COL INDEX="1">
					<NOTE_COUNT>1</NOTE_COUNT>
					<CONTENT>g</CONTENT>
				</COL>
				<COL INDEX="2">
					<NOTE_COUNT>1</NOTE_COUNT>
					<CONTENT>M</CONTENT>
				</COL>
				<COL INDEX="3">
					<NOTE_COUNT>1</NOTE_COUNT>
					<CONTENT>p</CONTENT>
				</COL>
				<COL INDEX="4">
					<NOTE_COUNT>1</NOTE_COUNT>
					<CONTENT>M</CONTENT>
				</COL>
			</ROW>
		</LINE>
	</LINES>
</SHEET>
\end{lstlisting}
    \caption{}
    \label{fig:part-2}
    \end{subfigure}
    \caption{Sheet part of the XML file (b) and (c) depicting part of the original music sheet (a). }
\end{figure}

\textbf{Sheet}, which is based on the 2D matrix model \textit{Swaralipi} \cite{misra2016swaralipi}, specifies the layout of the music sheet and the placement of the notation symbols. As a result, it replicates the entirety of the contents as a rectangular row-column arrangement. Even though we haven't yet transcribed the beat markings and lyrics, the model has the provision to include them in the future. The format cleverly transforms row and column models into helpful tags that make it easier to develop various applications, such as real-time note playback, producing music sheets, and retrieving score data. 
For example, part of the first line of the original composition (shown in Figure \ref{fig:sheet}) has been converted into the sheet part (shown in Figure \ref{fig:part-1} and \ref{fig:part-2}).

    

\section{Applications of the Dataset}

\begin{figure}
    \centering
    \includegraphics[scale=0.7]{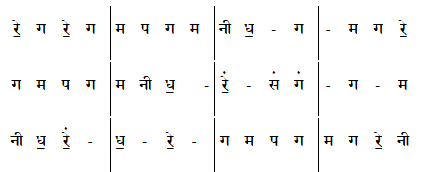}
    \caption{Music-sheet web visualization using Ome Swarlipi}
    \label{fig:visualization}
\end{figure}

\textbf{Visualization of Music-sheets:} One of the primary applications of any music dataset is to visualize it or render it using a notation system in which it is preserved. We have encountered several difficulties in visualizing the composition in the Bhatkhande notation system since there is no standard font system for rendering Bhatkhande music symbols in any language script. The closest rendition we have found is the \textit{Ome Swarlipi} \cite{omeswarlipi} system which is a compact version of the Bhatkhande notation system and easy to use. In order to visualize in HTML format, the system provides the necessary styling information to render it in \textit{Devanagari} script. Therefore the pre-processing step for this application is a converter that takes an XML file as a standalone composition and transforms it into equivalent HTML with the \textit{Ome Swarlipi} rendition of the score. The source code of the converter has been given in the \href{https://github.com/cmisra/Sangeet}{online repository link} and the corresponding rendition is shown in Figure \ref{fig:visualization}.

\textbf{Query and Retrieval of Musical Information:} This is the application where we can appreciate the power of XML as a means to build the music dataset. XML has brought with it a number of tools and technologies to efficiently process the information contained inside it. For the present application, we have used two tools, namely \textit{XPath} and \textit{XQuery}. XPath, the XML Path Language, uses path expressions to parse through the elements and attributes of an XML document and select node elements to extract the contents inside it. This language is also used in another query language XQuery to query an XML database and retrieve required information from it much like the SQL that does the same on a relational database.

\begin{figure}[!ht]
\centering
    \begin{subfigure}{0.48\textwidth}
\begin{lstlisting}[language=XQuery, label=lst:raag]
(: List of compositions having Meend :)
for $songs in collection ("Bhatkhande-Database")//swarlipi
let $title := $songs/INFO/TITLE/text()
let $contents := $songs/SHEET/LINES/LINE/ROW/COL/CONTENT/text()
let $notes := (for $song in $songs
return $song/SHEET/LINES/LINE/ROW/COL/CONTENT/text())
return if (contains(string-join($notes , ""),"q")) then
$title
\end{lstlisting}
    \caption{}
    \label{fig:query-1}
    \end{subfigure}
    \hfill
    \begin{subfigure}{0.49\textwidth}
\begin{lstlisting}[language=XQuery, label=lst:raag]
(: List of compositions having a particular Arohana subsequence :)
for $songs in collection ("Bhatkhande-Database")//swarlipi
let $title := $songs/INFO/TITLE/text()
let $aroha := $songs/RAAG/AROHANA/text()
return if (contains($aroha, "s-R-g")) then
$title
\end{lstlisting}
    \label{fig:query-1}
    \caption{}
    \label{fig:query-2}
    \end{subfigure} \\
    \begin{subfigure}{\textwidth}
\begin{lstlisting}[language=XQuery, label=lst:raag]
(:  Note frequency distribution of each composition :)
for $song in collection("Bhatkhande-Database")//swarlipi
let $raag := $song/RAAG/RAAG_NAME/text()
let $contents := $song/SHEET/LINES/LINE/ROW/COL/CONTENT/text()
let $joined_str := string-join(data($contents), ',')
let $joined_str := replace($joined_str, "<sup>|</sup>|@|u|l|\)|\(|-|,|\s+", "")
let $notes := (115,82,114,71,103,109,77,112,68,100,78,110)
let $code_points := string-to-codepoints($joined_str)
let $result := (for $i in $notes
    return count(index-of($code_points, $i))  )
let $result := normalize-space(string-join($result, ","))
return $result
\end{lstlisting}
    \caption{}
    \label{fig:query-3}
    \end{subfigure}
    \caption{XQuery to retrieve the (a) list of compositions having Meend, (b) List of compositions having a
particular \textit{Arohana} subsequence and (c) Note frequency distribution of each composition}
    \label{fig:xquery}
\end{figure}

The \textit{preprocessing} stage for this application is to create an XML database created from the XML documents. We have used \textit{BaseX} database engine to create the database from our dataset and XQuery to efficiently and easily perform complex queries and retrieve information from it and therefore, can be extremely useful for data-intensive complex web applications. This also provides a single-point query and retrieval system, as opposed to the current search and retrieval platforms \cite{tagoreweb,nltr} used for querying and browsing musical data. Figure \ref{fig:xquery} provides a few interesting and complex queries that satisfy the fine-grained information needs of the user. For example query \ref{fig:query-3} can be used to generate dataset for raag classification as described in the following section.

\begin{table}[!ht]
    \centering
    \begin{tabular}{|c|c|c|c|c|}
        \hline
        \multicolumn{5}{|c|}{\textbf{Accuracy Score of Classification Models}} \\
        \hline
        \textbf{Logistic Regression}  & \multicolumn{3}{|c|}{\textbf{K-Nearest-Neighbors (KNN)}} & \textbf{Decision Tree} \\
        \hline
         & $k=3$ & $k=5$ & $k=7$ & \\
         \hline
         0.9143 & 0.9714 & 0.9428 & 0.9428 & 0.9714 \\
         \hline
    \end{tabular}
    \caption{Performance measure of Logistic Regression, K-Nearest Neighbors with varying values of $k$, and Decision Tree. The dataset is divided into 70:30 as training and test set to calculate the accuracy score of different classification models.}
    \label{tab:ml}
\end{table}

\textbf{Raag prediction through Machine Learning:} This application refers to the musicological analysis of various musical components present in Hindustani Sangeet. It covers statistical and structural analysis, data mining, and inference using machine learning and deep learning techniques. As an example of the application, we apply machine learning techniques on the dataset for the task of raag prediction. The \textit{preprocessing} step for raag prediction is to convert the XML dataset into a tabular data-frame containing a number of features and a target variable. For raag prediction, we take features as the frequencies of individual notes and the corresponding raag as a target variable for any composition. Instead of taking the note-frequency distribution of $36$ notes for a composition spanning across three octaves, we merge the notes to obtain the frequency distribution of $12$ notes. Since, the positions of the notes of the \textit{Arohana} and \textit{Avarohana} of any particular composition in different octaves do not affect the raag of the composition, we map corresponding notes of three octaves and make a sum of frequencies of corresponding notes to obtain $12$ note-frequency distribution (can be obtained from \ref{fig:query-3} given in \href{https://github.com/cmisra/Sangeet}{GitHub}). Table \ref{tab:ml} shows the measure of performance of different machine learning techniques for raag prediction for our dataset. We have transformed our dataset into a three-class classification problem by taking the three most frequent raags i.e. \textit{Bhairav}, \textit{Todi}, and \textit{Poorvi}, and applied the different classification models to generate the accuracy scores. Since each classifier examined shows high accuracy score the dataset can be considered as a robust dataset for raag classification. Table \ref{tab:ml} shows that KNN with $k=3$ and decision tree classifier gives better accuracy scores than the logistic regression model.

\section{Conclusions and Future Works}
\label{sec:conclusions}
This paper presents SANGEET, a Hindustani Sangeet dataset based on XML to provide easy and efficient access to a music corpora to perform various applications including music visualization, MIR, and Raag prediction using machine learning techniques. Backed by a robust music-sheet framework and a structured XSD, SANGEET provides a comprehensive repository for rich musical information to be shared seamlessly across applications. We have shown that SANGEET is quite efficient for accessing and transforming musical data into a format suitable for various musical applications. Our future objective is to extend SANGEET with the compositions of Bhatkhande's other five volumes of \textit{Kramik Pustak Malika} and update the structure of the XML dataset with taal markings and lyric information. This will provide better music-sheet rendition and richer queries to fulfill the user's information needs.

\bibliographystyle{splncs} 
\bibliography{cmmr2023}

\end{document}